\documentclass[9pt,twocolumn,twoside]{osajnl}
\journal{ol} % Choose journal (ao, aop, josaa, josab, ol, pr)

\usepackage{soul}
\usepackage{color}

\setboolean{shortarticle}{true}
\title{Dual-comb correction with spectrally broadened fiber lasers}

\author[1]{Philippe Guay}
\author[1]{Alex Tourigny-Plante}
\author[1]{Nicolas Bourbeau Hébert}
\author[1]{Vincent Michaud-Belleau}
\author[1]{Steeve Larouche}
\author[1]{Khaoula Fdil}
\author[1,*]{Jérôme Genest}

\affil[1]{Centre d'optique, photonique et laser, Universit\'{e} Laval, Qu\'{e}bec, Qu\'{e}bec G1V 0A6, Canada}

\affil[*]{Corresponding author: jerome.genest@copl.ulaval.ca}

\begin{abstract}
The phase information provided by the beat note between frequency combs and two continuous-wave lasers is used to extrapolate the phase evolution of comb modes found in a spectral region obtained via nonlinear broadening. This thereafter enables using interferogram self-correction to fully retrieve the coherence of a dual-comb beat note between two independent fiber lasers. This approach allows to forego the $f-2f$ self-referencing of both combs, which is a significant simplification. Broadband near-infrared methane spectroscopy has been conducted as a demonstration of the simplified system's preserved performance. 
\end{abstract}

\setboolean{displaycopyright}{true}

\begin{document}

\maketitle

Dual-comb interferometers, which can be used as high precision spectrometers, have been widely developed for atmospheric sensing \cite{YCA19, FLE16,RIE14}. For instance, greenhouse gases such as methane have been extensively studied in controlled environments \cite{FDI19,GUA19,CHE19,OKU15}. However, comb-based gas sensing outside the laboratory brings a new set of challenges \cite{SIN14}. Recently, there has been field demonstrations of methane detection addressing those challenges \cite{COB18,RIE14}, but these dual-comb systems were based on fully locked combs and self-referenced combs to perform spectroscopy. As the need for portable instrument requires the system to be compact and cheap, further simplifications would be beneficial.

Self-correction algorithms have been proven to be effective with free-running dual-comb lasers originating from the same laser source \cite{GUA18,FDI19} or in the case of highly mutually coherent lasers \cite{HEB18,ZHA16}. The range of operation of our self-correction algorithm is now extended to independent lasers with only minimal active stabilization. 

In this Letter, we present simplifications to a system used for methane sensing \cite{SIN15,COB18}. Reduction of the system's complexity is achieved through the introduction of a second continuous-wave (CW) laser to the experimental setup and the use of a self-correcting algorithm \cite{HEB17}. The $f-2f$ interferometer originally used to stabilize the carrier-enveloppe offset (CEO) frequency is proven to become unnecessary as phase drifts are corrected in post-processing. As an additional simplification, the "fast" piezo-electric transducer (PZT) used to stabilize rapid fluctuations of the laser's repetition rate is dismissed, leaving only the "slow" PZT to minimally stabilize the comb. To demonstrate the performance of the simplified spectrometer, the lasers are amplified and broadened in highly nonlinear fiber (HNLF) up to 1700 nm to interrogate methane's 2$\nu_3$ overtone. The absorption spectrum is measured and compared for the actively locked and the minimally stabilized cases. A comparison of a spectral line to HITRAN database is also provided to assess the validity of the measurement.

\begin{figure}[h]
\centering
\includegraphics[width=\linewidth]{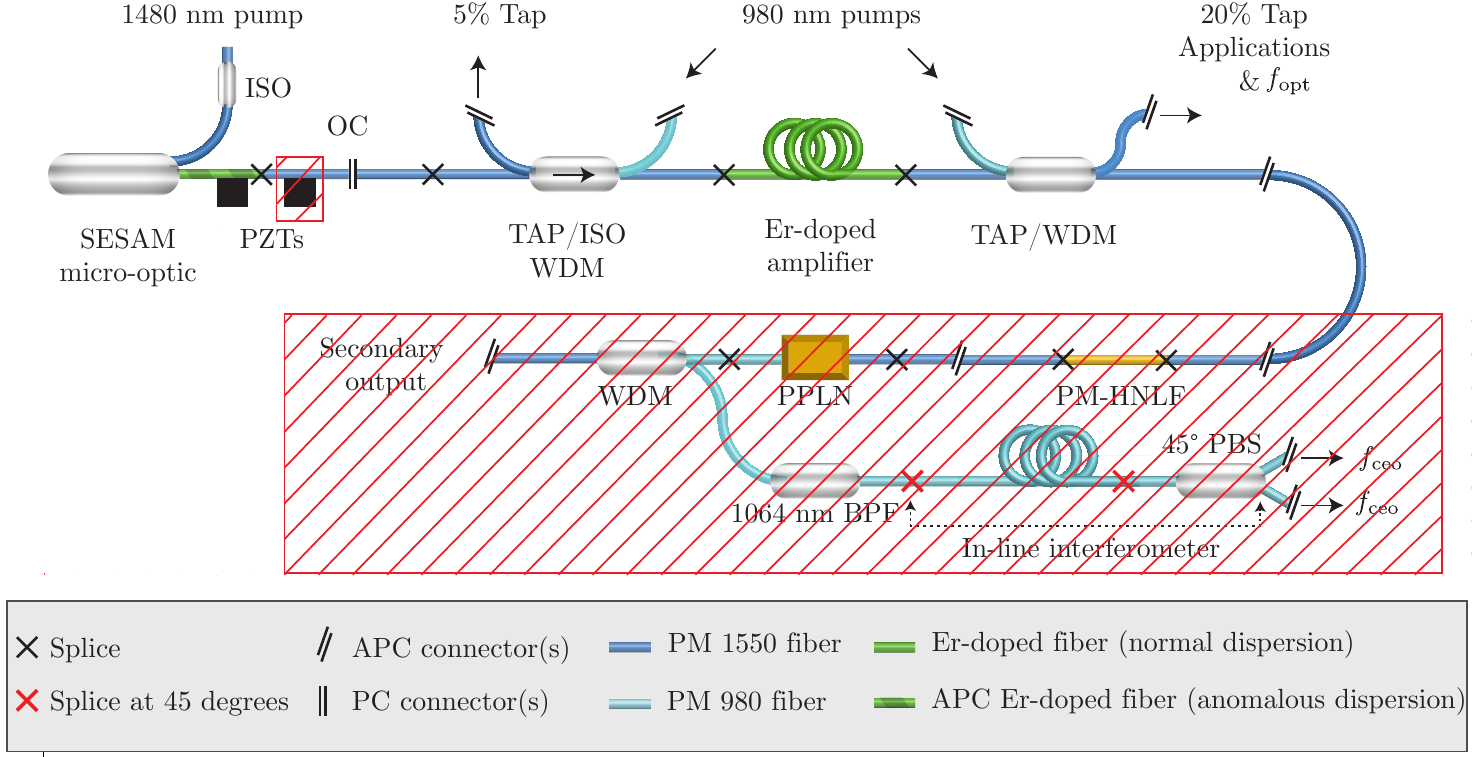}
\caption{Schematic of a frequency comb reproduced from \cite{SIN15} where components substituted by a second CW laser and a self-correction algorithm are indicated by the red lined areas. SESAM, semiconductor saturable absorber mirror; PZT, piezo-electric transducer; OC, output coupler; ISO, isolator;  WDM, wavelength division multiplexer; PM-HNLF, polarization-maintaining highly nonlinear fiber; PPLN, periodically poled lithium niobate; BPF, bandpass filter; PBS, polarization beam splitter.  }
\label{fig:exp}
\end{figure}

\begin{figure*}[h]
\centering
\includegraphics[width=\linewidth]{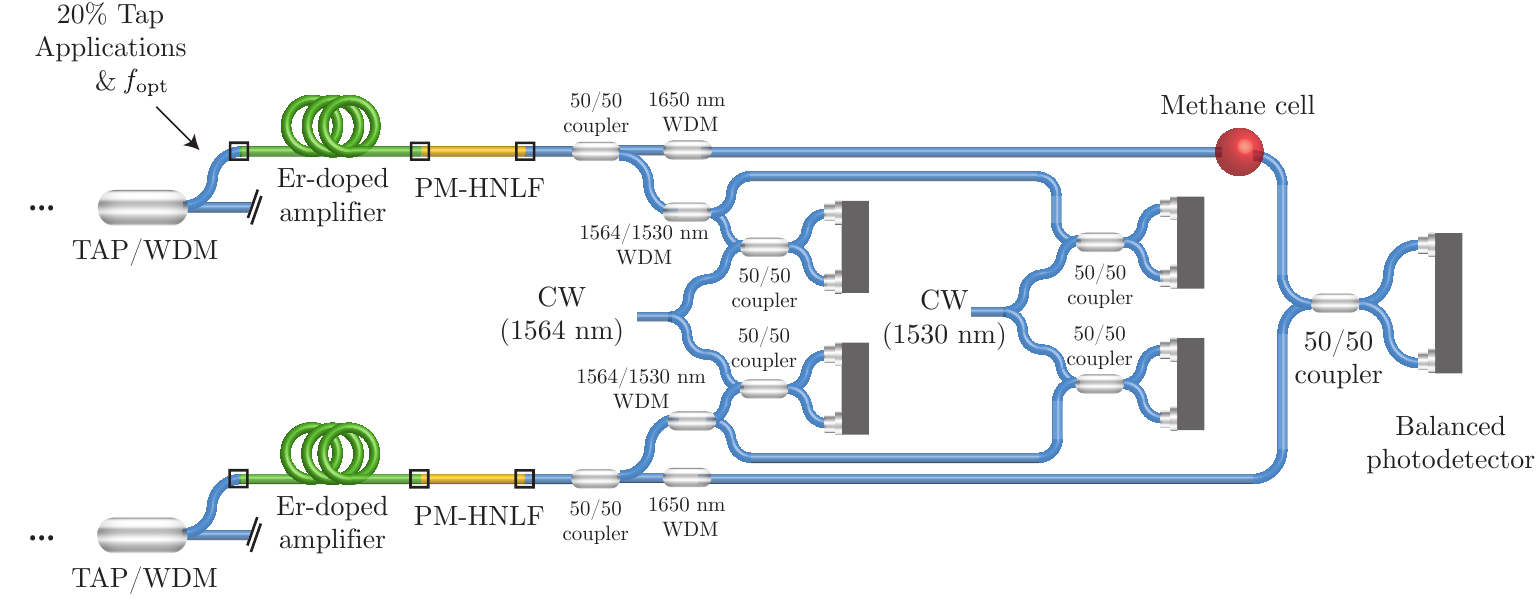}
\caption{Experimental setup for the dual-comb experiment.WDM, wavelength division multiplexer; PM-HNLF, polarization-maintaining highly nonlinear fiber; CW, continuous-wave laser.}
\label{fig:exp2}
\end{figure*}

The comb sources used are self-referenced fiber lasers based on the design fully described in \cite{SIN15}. Fig. \ref{fig:exp} shows one such source. Here, two frequency combs with repetition rate $f_r~\approx~160$ MHz and repetition rate difference $\Delta f_r~=~828$ Hz are used. In the figure, the self-referencing section has been shaded to illustrate that it is no longer needed to produce coherent interferograms (IGMs) when the processing steps proposed in this paper are followed. This significantly reduces the hardware complexity of both sources since the PPLNs are no longer needed and the requirements are greatly relaxed for the HNLF as coverage for an octave  is no longer needed. The spectral broadening can thus be optimized for spectroscopy alone.

To perform spectroscopy, the frequency combs are spectrally broadened up to 1700 nm using amplifiers and highly nonlinear fiber (OFS HNLF-PM). This second pair of amplifiers would not be required if the 20\% TAP/WDMs were replaced by simple WDMs in order to give access to the full power generated by the first pair of amplifiers. Each comb is optically filtered with a bandpass filter centered on 1650~nm (Advance Fiber Resources).

The dual-comb experimental setup is shown in Fig. \ref{fig:exp2}. One of the comb interrogates a 5.5-cm long cell filled with methane at atmospheric pressure 740$\pm$10\%~Torr (Wavelength References). The two combs are recombined with a 50/50 coupler and mixed to produce an RF comb. Only one interferometer is used for spectroscopy since no calibration of the absorption spectrum is performed for this demonstration. The data is acquired on a GaGe acquisition board (CSE8389) with a sampling rate of 125 MS/s. 

The amplified and broadened combs are split twice and mixed with two CW lasers. The 1564 nm beat signal is used to provide minimal active stabilization. A longitudinal mode of each comb at 1564~nm is mixed with the laser (RIO Planex RIO0095-3-15-1) and the beat note is fed to an FPGA running our open-source phase-locked loop software \cite{TOU18} to actively correct phase drifts of the comb at 1564~nm with a "slow" (10 kHz bandwidth) PZT controller (or actuator). The "fast" PZT controller (1 MHz bandwidth) that was in the initial design \cite{SIN15} has been removed since it introduces fast noise around its resonance frequency that degrades the quality of the dual-comb interference. To compensate the additional fluctuations not corrected by the PZT controller, the beat note fed to the FPGA is also sent to the acquisition board and a phase correction is performed in post-processing before using the self-correction algorithm.

% The first idea was that the stabilization and correction of the mutual phase between the combs at 1564~nm would allow using interferogram self-correction of the beat note around 1650~nm. In principle, with variations removed, only an interferogram resampling, with a fixed-point at 1564~nm is needed and this operation is not subject to phase unwrapping issues.  However, significant fluctuations of the repetitions rates in a band larger than $\Delta f_r/2$ prohibit success with self-sufficient algorithms \cite{HEB19}. Because of that, the mutual coherence between the combs could not be retrieved, teeth were smeared and a resolution loss was observed on the methane lines. This is shown in Fig. \ref{fig:methane_smeared} which compares the case of the phase correction with one CW laser with the fully stabilized case. The line R(3) appears broadened by improperly corrected interferograms.

\begin{figure}[h]
\centering
\includegraphics[width=\linewidth]{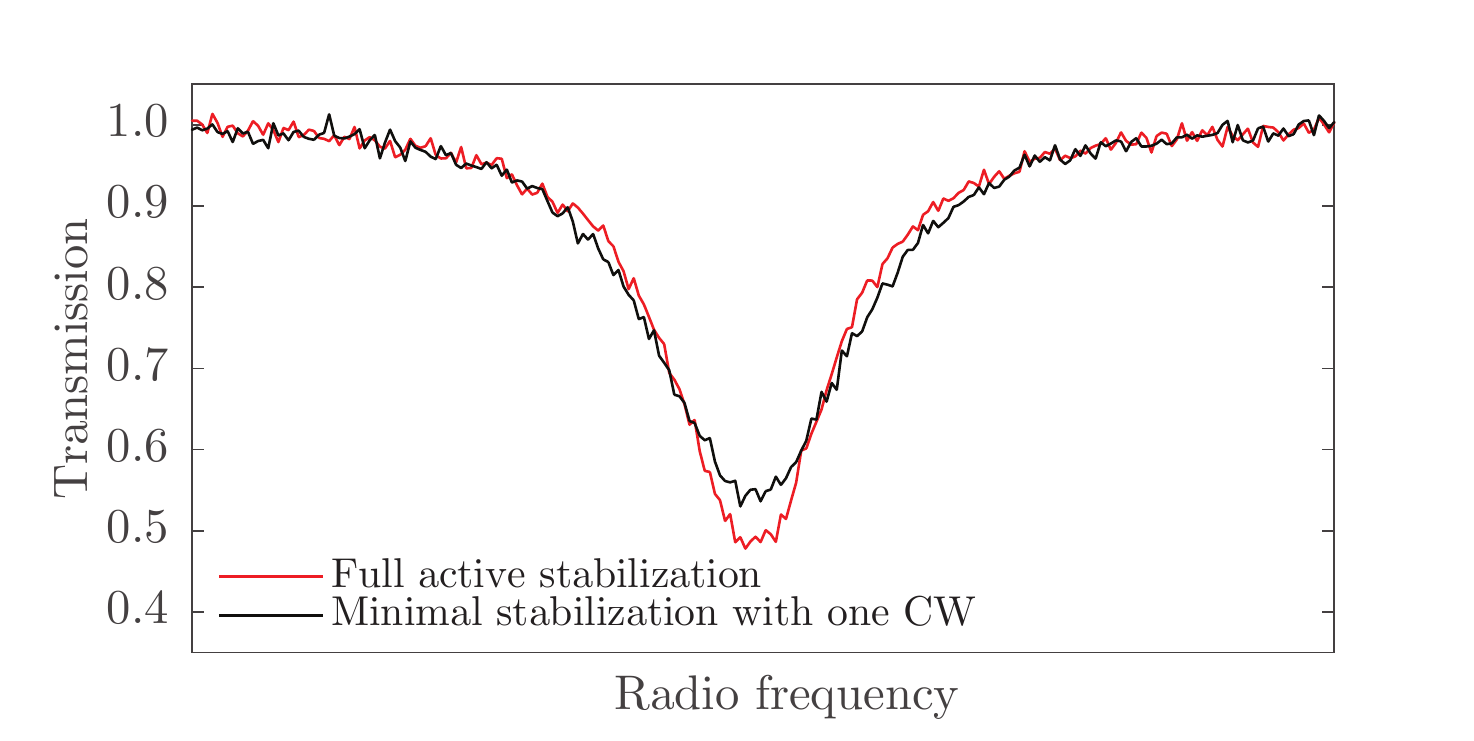} 
\caption{Transmission spectrum in the 2$\nu_3$ R(3) methane manifold region for the case of a phase correction with one CW laser and the fully stabilized case.}
\label{fig:methane_smeared}
\end{figure}

The first idea was that the stabilization and correction of the mutual phase between the combs at 1564~nm would allow using the interferogram self-correction algorithm for the 1630-1670~nm region. In principle, with phase variations corrected at 1564~nm, only fluctuations of the repetition rate difference remain. These can be corrected with a resampling operation without phase wrapping issues. However, significant fluctuations of the repetitions rates with a bandwidth larger than $\Delta f_r/2$ prohibit success with self-sufficient algorithms \cite{HEB19}. For that reason, the mutual coherence between the combs could not be retrieved, teeth were smeared, and a resolution loss was observed on the methane lines. This is shown in Fig. \ref{fig:methane_smeared} which compares the case of the phase correction with one CW laser at 1564~nm followed by the self-correction with the  the fully stabilized case. The line R(3) appears broadened because of an inadequate correction of the IGMs' repetition rate.

% To solve this problem of uncorrected fluctuations, a CW laser operating at 1530~nm (RIO Planex RIO0095-3-59-1) is introduced to measure the combs' phase at a second location. An interferogram software post-correction using two CW lasers as anchors to pin the beatnote spectrum at two locations from either side of a band of interest has already been demonstrated \cite{DES10}. This technique is reproduced here with two lasers at 1530~nm and 1564~nm. The idea behind the use of two lasers is to provide an estimate of the CEO frequency $\phi_\text{ceo}$ and repetition rate difference phase $\phi_{dfr}$ of the RF comb. Knowing these two quantities allows to calculate the phase at a desired location. Here, the phase at 1650~nm will be extrapolated to pre-correct the interferogram such that self-correction can afterwards successfully retrieve the mutual comb coherence. One shall note that the pre-correction does not need to be perfect. It just needs to meet the self-correctability criteria \cite{HEB19}.  The phase at 1640 nm can be estimated with the following equations, where $k$, $l$ and $m$ represent the mode number at 1530~nm, 1564~nm and 1640~nm respectively (Equations \ref{phi_dfr}).

To solve this problem of uncorrected fluctuations, a CW laser operating at 1530~nm (RIO Planex RIO0095-3-59-1) is introduced to measure the combs' relative phase at a second spectra location. The idea behind the use of two lasers is to provide an estimate of the CEO frequency $\phi_\text{ceo}$ and repetition rate difference phase $\phi_{dfr}$ of the RF comb \cite{DES10}. Knowing these two quantities allows to calculate the phase at a desired frequency. Here, the phase at 1640~nm is extrapolated to pre-correct the interferograms such that self-correction can afterwards successfully retrieve the mutual comb coherence. One shall note that the pre-correction does not need to be perfect. It only needs to meet the self-correctability criterion \cite{HEB19}.  The phase at 1640 nm can be estimated with the following equations, where $k$, $l$ and $m$ represent the mode number at 1530~nm, 1564~nm and 1640~nm respectively (Equations \ref{phi_dfr}).

% An interferogram software post-correction using two CW lasers as anchors to pin the beatnote spectrum at two locations from either side of a band of interest has already been demonstrated \cite{DES10}. This technique is reproduced here with two lasers at 1530~nm and 1564~nm. 

%These two quantities can be determined from any reference point, but it is advantageous to use 1564 nm as the location from which there is no repetition rate drifts since it is closer to the region of interest. Accordingly, the phase offset is equal to the measured phase at 1564 nm (Equation \ref{phi_CEO}) and the repetition rate phase become the phase difference between the measured phase at 1530 nm and 1564 nm divided the difference of the mode number $m$ and $k$ at 1530 nm and 1564 nm respectively (Equation \ref{phi_dfr}). 

\begin{subequations}
\label{phi_dfr}
\begin{align}
    \phi_\text{$df_r$} &= \frac{\phi_\text{1530 nm}-\phi_\text{1564 nm}}{k-l} \\
    \phi_\text{1640} & = \phi_\text{1564}+(m-l)\phi_\text{$df_r$} \label{phi_1640}
\end{align}
\end{subequations}

% The calculation of the CEO frequency, and the repetition rate difference phase allows to calculate the phase at the desired frequency. The number of mode $l$ at 1640 nm is used to calculate the phase in the band of interest.

% \begin{align}
%     \phi_{1640} = \phi_\text{offset} + (l-m)\phi_\text{d$f_r$}
% \end{align}

After the pre-correction, the self-correction algorithm previously reported in \cite{HEB17} extracts information from the IGMs themselves to perform phase correction for the CEO frequency and a resampling according to the repetition rate difference. Since the phase fluctuations occurring in the interferometer are not taken into account as they are not measured by the CW lasers, a self-correction is still required to properly phase-correct the IGMs. 

% \hl{in the following paragraph, we have to sort who did what JR, JD and the diffenrece between their phase-correction and the self-correction.}
% A complete correction of the spectrum at 1640 nm with the two CW lasers technique could have been done such as in \cite{ROY12}, but the phase fluctuations occurring in the interferometers are not taken into account as they are not measured by the CW lasers unless a subsequent cross-correlation of the IGMs is performed as in \cite{DES10}. It thus justifies the need to used a self-correction algorithm that account for "out of loop" phase noise. The algorithm's correction bandwidth is however limited to half the rate of the IGMs ($\Delta f_r /2$) since a phase and a timing estimate are measured once per cycle. The combination of both techniques ensures that the phase noise "out of loop" is corrected with the self-correction algorithm (within $\Delta f_r /2$ bandwidth) and that the phase noise "in-loop" is corrected with the CW lasers measurement (within $f_r$ bandwidth).

\begin{figure}[h]
\centering
\includegraphics[width=\linewidth]{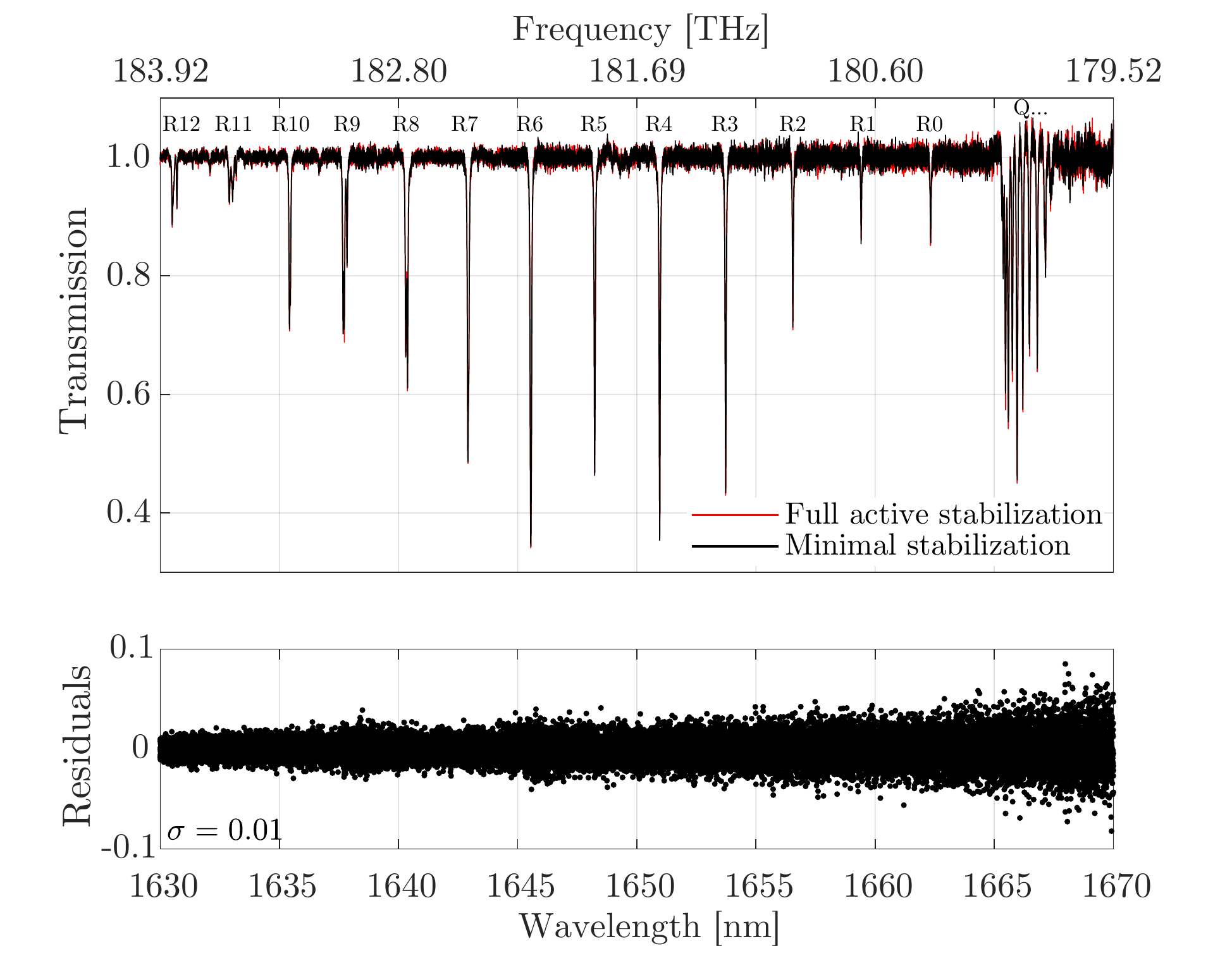}
\caption{Transmission spectrum of methane (R and Q branches) for the cases of a fully locked (red) and minimally stabilized frequency combs (black) where no distinction is visible within 1 \%. }
\label{fig:methane}
\end{figure}

The transmission spectrum of methane is presented in Fig. \ref{fig:methane}. The R and Q branches of the 2$\nu_3$ overtone are shown for the case of a self-referenced system  (full active stabilization) and for the case of the simplified system with two CW lasers (minimal stabilization). Reasonable correspondence is observed as no notable difference is visible between the two cases. The residuals confirm that the two curves are the same within 1\%, which is the standard deviation of the residuals that are limited by the measurement noise. The increasing variance of the residuals towards longer wavelengths is explained by the lower optical signal-to-noise ratio as wavelength increases. 

% Nécessaire?
% Even if the phase of the complex absorption spectrum was measured by the placing the gas cell before the interferometer, it is not shown here as unwrapping issues have arisen due to the mismatched delays between the arms of the interferometers. A delay line compensating the delay introduced by the gas cell would have corrected the problem. However, a comparison of the modulus of the complex transmission spectrum is sufficient for the needs of this demonstration. Only one arm of the interferometer interrogates the gas cell to take advantage of the balanced photodetection.  

\begin{figure}[h]
\centering
\includegraphics[width=\linewidth]{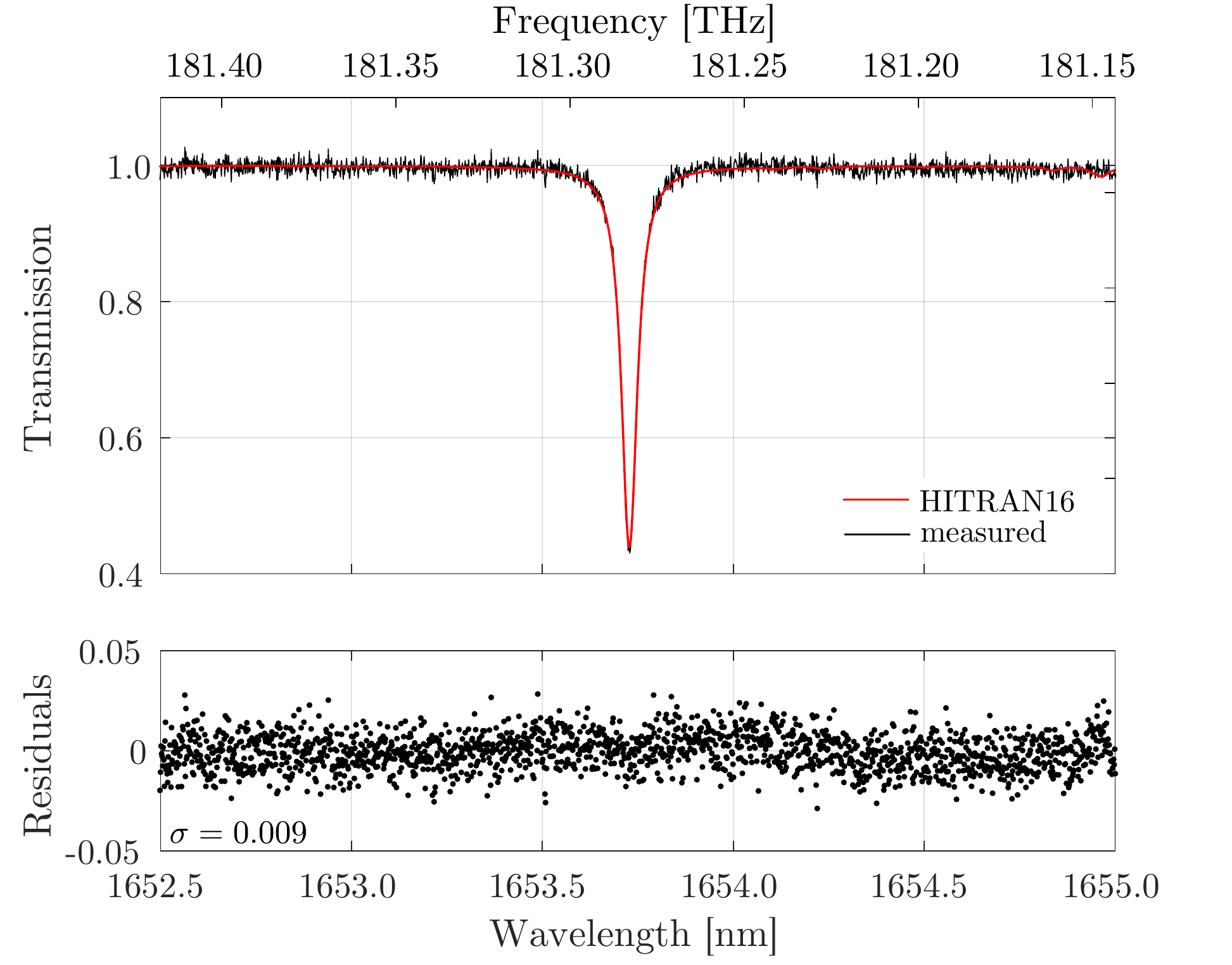}
\caption{Transmission spectrum (top) in the 2$\nu_3$ R(3) methane manifold region for the experimental data without CEO servo-loop (black) and as modeled with Voigt line shapes computed with parameters from HITRAN 2016 database (red) and the corresponding residuals (bottom). Standard deviation of residuals $\sigma$ is given on the bottom panel.}
\label{fig:methaneHITRAN}
\end{figure}

Sixth-order piece-wise polynomial functions, where each function spans approx 4~nm around each absorption line, have been removed from the baseline of the data to produce a flat-looking spectrum allowing better comparison between presented scenarios. The frequency axis of Fig. \ref{fig:methane} has been determined from an HITRAN fit that is shown in Fig. \ref{fig:methaneHITRAN} and detailed below. As the dual-comb experiment performed here did not provide an absolute frequency reference, an absorption line of methane is used to provide an estimate of the absolute frequency of the measurement.

To assess the validity of the measurement, the R(3) manifold of methane in the 2$\nu_3$ band has been fitted to a modeled spectrum using a complex Voigt profile. Data from HITRAN 2016 has been used. Excellent correspondence is observed between the experimental data and the theoretical model, which confirms that the performance of the spectrometer is unaltered by the suggested simplifications.

The fit was performed with an optimization function that minimized the least-square error on the residuals where the pressure of the cell, its length, and its temperature were free parameters. The absolute frequency of the acquired data, the optical point spacing ($f_r$ ), and the coefficients of a global fourth-order polynomial function were also parameters of the optimization.  A pressure of 92 kPa, a cell length of 5.3 cm and temperature of 21 $^\text{o}$C were found, matching the uncertainty bounds given by the manufacturer.

%\begin{figure}[htbp]
%\centering
%\includegraphics[width=\linewidth]{sample}
%\caption{False-color image, where each pixel is assigned to one of seven reference spectra.}
%\label{fig:false-color}
%\end{figure}

In conclusion, simplifications to an existing dual-comb interferometer have been demonstrated. The simplifications greatly reduce the complexity of the instrument and its cost, thus improving its quality as a field-deployable instrument. Nonlinear broadening of the laser's spectrum has been realized to perform near-infrared broadband spectroscopy of methane. Branches R and Q of methane in the $2\nu_3$ band were measured with and without the self-referencing interferometer to assess the performance of the simplified system.

\section*{Funding Information}
Natural Sciences and Engineering Research Council of Canada (NSERC); Fonds de Recherche du Québec-Nature et Technologies (FRQNT).

\section*{Acknowledgement}
The authors thank Ian Coddington at NIST for providing the dual comb system.

\section*{Disclosures}
The authors declare no conflicts of interest.

% Bibliography
\bibliography{sample}

% Full bibliography added automatically for Optics Letters submissions; the following line will simply be ignored if submitting to other journals.
% Note that this extra page will not count against page length
\bibliographyfullrefs{sample}

\end{document}